\documentclass[doublecol]{epl2} 

\title{A novel non-Fermi-liquid  state in the iron-pnictide FeCrAs}

\author{W. Wu\inst{1} \and A. McCollam\inst{1} \and I.\ Swainson\inst{2} 
\and P.M.C. Rourke\inst{1} \and D.G. Rancourt\inst{3} \and S.R. Julian\inst{1}}

\institute{                    
  \inst{1} Department of Physics, University of Toronto, 60 St. George Street, Toronto, Canada M5S 1A7\\
\inst{2} Canadian Neutron Beam Centre, National Research Council Canada, Building 459, Chalk River Laboratories, Chalk River, Ontario, Canada K0J IJ0

\inst{3} Department of Physics, University of Ottawa, Ottawa, Ontario, Canada K1N 6N5\\
}


\abstract{
We report transport and thermodynamic properties of stoichiometric 
single crystals 
of the hexagonal iron-pnictide FeCrAs. 
The in-plane resistivity shows an unusual ``non-metallic" dependence on 
temperature $T$, 
rising continuously with decreasing $T$ from  $\sim$800~K to
below 100 mK. 
The $c$-axis resistivity is similar, except for a sharp
drop upon entry into an antiferromagnetic state at $T_N \sim
125$~K. Below 10~K the resistivity follows a 
non-Fermi-liquid power law, 
$\rho (T)=\rho_0 -AT^x$ with $x<1$, while the specific heat shows Fermi
liquid behaviour
with a large Sommerfeld coefficient, $ \gamma \sim 30 {\rm\  mJ/mol\,K^2} $. 
The high temperature properties are 
reminiscent of those of the parent compounds of 
the new layered iron-pnictide superconductors, 
however the \(T \rightarrow 0{\rm ~K}\)  
properties suggest a new class of non-Fermi liquid. 
}

\begin{document}

\maketitle
\bibliographystyle{eplbib}

\section{Introduction}

Metallic states that violate Landau's Fermi liquid paradigm 
appear in 
many strongly correlated electron systems such as 
doped cuprates, 
quantum critical metals, and disordered Kondo lattices 
\cite{Hussey08,VonLohneysen07,Stewart01}.
The new layered iron-pnictide high temperature superconductors 
also show non-Fermi-liquid behaviour, 
most notably in the undoped parent compounds which have 
a high, roughly constant resistivity 
at temperatures above a magnetic 
spin-density-wave transition that typically occurs 
around \( T_{SDW} \sim \rm 150\ K \) 
(see, for example, \cite{Kamihara,Ren, Rotter,Sasmal}). 
For \( T < T_{SDW} \), however, this incoherent charge transport 
gives way to a 
rapidly falling (i.e.\ ``metallic'') resistivity.

The basic structural units of the new iron-pnictide superconductors  
are layers 
in which iron atoms are tetrahedrally coordinated by arsenic. 
Some theoretical approaches to the incoherent $T > T_{SDW}$ 
state in these systems suggest that the physics is 
different from that of the cuprates, not least because 
this tetrahedral coordination produces only a small crystal-field 
splitting of the Fe $3d$-shell. 
Sawatzky et al.\ \cite{Sawatzky} 
argue that the electronic structure is dominated by 
narrow iron $3d$ bands coupled to a highly polarizable arsenic 
environment. 
Dynamical mean field theory (DMFT) treatments 
\cite{Shorikov,Haule,Laad} 
find an incoherent metallic state that arises from 
strong on-site inter-orbital interactions in the 
$3d$-shell that perhaps puts some \cite{Shorikov} 
or all \cite{Haule} of the weakly crystal field split 
$3d$-orbitals close to a Mott transition.
The local physics of this so-called ``bad semiconducting'' state 
may represent a qualitatively new kind of 
incoherent metal.

In this paper we present magnetic, thermodynamic and 
transport properties of 
a ternary iron-arsenide, FeCrAs,  
whose behaviour shows intriguing 
similarities to the parent compounds of the layered iron-pnictide superconductors, 
including an antiferromagnetic transition in a similar temperature range 
($T_N \sim 125$~K). 
However there are key differences: 
firstly, the crystal structure is hexagonal as opposed to tetragonal; 
secondly, 
although the iron atoms are tetrahedrally coordinated by 
arsenic, FeCrAs is three-dimensional (it does not have insulating layers); 
thirdly,   as shown below,  
in FeCrAs the magnetism for \( T < T_{N} \) resides 
primarily on the Cr sites, whereas in the layered iron-pnictides  
the magnetism is on the iron sites; 
and finally, the 
behaviour of the 
electrical resistivity of FeCrAs is much more extreme 
in that it is ``non-metallic" 
over a huge temperature range -- 
the in-plane resistivity rises with decreasing temperature 
from $\sim$800~K to below 100 mK 
without any sign of saturation or  a gap at low temperature.  
Moreover, there is 
a profound and novel incompatibility  
at low temperature between the resistivity, 
which is non-Fermi-liquid with a sub-linear 
dependence on temperature, 
and the specific heat,  which  has a Fermi-liquid 
linear-in-\(T\) dependence 
as \( T \rightarrow 0\)~K.

\section{Experiment}

FeCrAs is a member of a large family of ternary 3$d$ transition metal
monopnictides \cite{Fruchart}, many  of which have a tetragonal crystal
structure and order antiferromagnetically above room temperature. 
FeCrAs, however, 
has the hexagonal $\rm {Fe_2P}$ structure, and generally members of
this family with this structure  order magnetically below room temperature, with
ferromagnetic or complex antiferromagnetic order (see e.g. \cite{Bacmann}).

Single crystals of FeCrAs  were  grown by melting   
stoichiometric quantities of high purity Fe, Cr and As 
following the recipe given in \cite{Katsuraki66}.
Several batches of crystals   were grown and  annealed at $900^{\circ}$C in
vacuum for five to ten days. Some batches were characterized by x-ray
powder diffraction, which showed that the samples  crystallized  in
the correct  structure,  with space group P$\bar{6}2m$, and no detectable
trace impurity phases. 
The crystals are shiny and metallic looking, but  brittle. 
Material from one batch was examined by
 powder neutron diffraction at temperatures 
down to 2.8~K. 
In addition to finding magnetic order at low temperature (discussed below) 
this showed that there is a high degree of site-order 
on the Fe and Cr sublattices.  
All measurements were performed
on single crystals except for the x-ray and neutron  diffraction.

Transport measurements used a conventional four terminal 
technique, either in a dilution refrigerator with a superconducting 
magnet for measurements below room temperature, or in a tube furnace 
under vacuum for high temperatures. 
In the latter measurements, the maximum temperature was 
limited by evaporation of the gold leads. 
Magnetic susceptibility measurements were done in a Quantum Design 
MPMS system, while specific heat measurements were done in
a Quantum Design PPMS system. 

\section{Results}

\begin{figure}
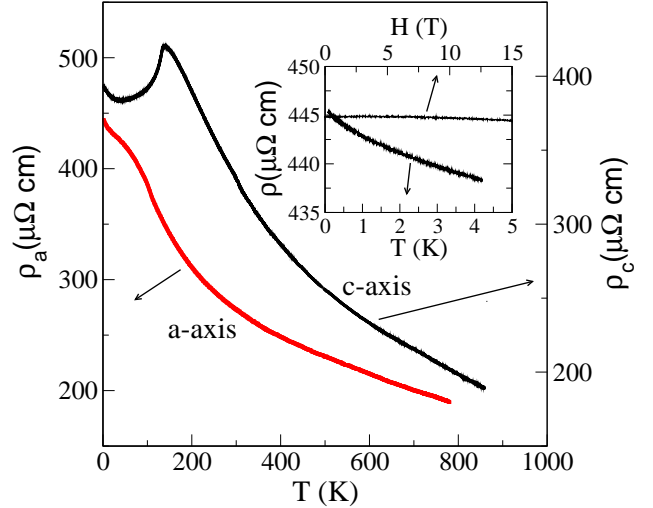

\onefigure[scale=0.36]{./Julian_Fig1.eps}
\caption{Resistivity vs.\ temperature for $a$-axis (red curve) and 
$c$-axis (black curve).  The inset shows the 
$a$-axis resistivity between  80 mK and 4~K (bottom and left axes), 
demonstrating that it does not 
saturate or show a gap-like structure down to the lowest temperature 
measured, together with 
the resistivity vs. field at 170~mK (upper and left axes), 
showing that the magnetoresistance is extremely weak .}
\label{fig-rhoVsT}
\end{figure}

The main panel in Fig. \ref{fig-rhoVsT} shows the temperature dependence of resistivity. 
The absolute
resistivity $\rho$ along the  $a$ and $c$ axes ranges from $\sim$200 to $\sim$500
$\rm {\mu\Omega~cm}$, which at high temperature at least 
is typical for a strongly correlated metal. 
There is only weak anisotropy in the resistivity, reflecting the absence of 
insulating layers in this material. 
The $a$-axis  resistivity \( \rho_a(T) \) shows a remarkable non-metallic 
behaviour, growing 
with decreasing temperature from $\sim$800~K down to at least 80 mK. 
An expanded
plot of the  $a$-axis resistivity at low temperature is shown in 
the Fig. \ref{fig-rhoVsT}
inset, where it can be seen that   
there is 
no sign of saturation or a gap that would
give rise to exponentially increasing resistivity as $T$ falls. 
Instead, \(\rho_a(T) \) has an exotic non-metallic and sub-linear in \(T\) 
behaviour as \(T \rightarrow 0{\rm ~K}\):  $ \rho_a(T) = \rho_{a,0} - A T^{0.60\pm0.05}$.  

The $c$-axis resistivity shows similar behaviour 
except that there is a very pronounced maximum, or
cusp, at  $\sim$125~K.  
Examined  closely, the in-plane resistivity also
shows  a slope change  at approximately the  same temperature.  
Below 10~K \(\rho_c(T) = \rho_{c,0} - A T^{0.70\pm0.05}\),  
which is again sub-linear in \(T\).
At high temperatures, we followed the  
negative  $d\rho/dT$ to nearly 900~K without finding any sign of saturation.
Both the temperature at which the cusp appears and
the magnitude of the drop in $\rho_c$ below this cusp have large
sample to sample variations, however, the low temperature power law, and
the resistivity above $\sim$150~K, are sample independent.

The inset of Fig. \ref{fig-rhoVsT} shows that the field dependence of $\rho$
is very weak, decreasing by $\sim 0.08\%$ from 0 to 15 Tesla at 170~mK.   We have also
measured the Hall effect both at room temperature and between 80-300 mK,
but find that the Hall signal is too small to be resolved.  Based on
the resolution of our devices, the low-temperature and room-temperature
Hall coefficients are estimated to be less than 4.5$\rm {\times 10^{-3}\
cm^3/C}$ and 3.2 $\rm{\times 10^{-3}\ cm^3/C}$, respectively.
This suggests that FeCrAs is a compensated metal with a scattering rate 
that is uniform over the Fermi surface.

\begin{figure}
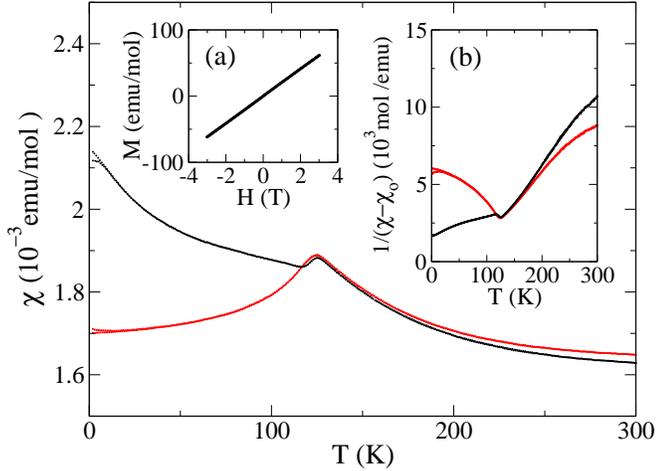
 
\onefigure[scale=0.34]{./Julian_Fig2.eps}
\caption{ Temperature dependence of the susceptibility $\chi$  for
fields along $a$-axis (red curve) and $c$-axis (black curve). 
A small difference between zero-field-cooled and field-cooled conditions 
is visible at the lowest temperatures.  
Inset: (a) the temperature dependence of $1/(\chi - \chi_0$) showing an 
absence of Curie-Weiss behaviour; (b) a hysteresis plot of 
the field dependence of the magnetization at 2~K.}
\label{fig-chiVsT}
\end{figure}

Fig. \ref{fig-chiVsT} shows the temperature dependence of the  magnetic
susceptibility $\chi$ for fields along $a$- and $c$-axis,  under both
field cooled (FC) and zero field cooled (ZFC) conditions.  For both
orientations, a clear  peak is observed at 125~K, below which $\chi_a$
and $\chi_c$ deviate from one another.  
In addition, at lower temperatures a weak difference appears between 
the  FC and ZFC susceptibilities, which is indicative of magnetic 
freezing as is observed in spin-glasses. 
In Fig. \ref{fig-chiVsT} this magnetic freezing sets in below 
7~K and 15~K for $H\parallel c$ and $H\parallel a$ respectively; 
however,  we have found that this
behaviour is very sample dependent, starting  as high as $\sim$45~K in
some crystals. The $\rho(T)$, $C(T)$ and $\chi(T)$ data shown in this paper
come from our ``best'' samples, that is, those with the lowest freezing
temperature. However, we emphasize that, aside from the size of the
drop in $\rho_c(T)$ below 125~K (which is largest in the ``best'' samples),
none of these quantities showed qualitatively different behaviour between
samples.  

For all samples,  no magnetic hysteresis was observed down to
2~K: the inset of Fig. \ref{fig-chiVsT} shows an example of the field dependence of the magnetization  at 2~K.  

The overall change
of the susceptibility is small in the whole temperature range observed,
that is, \( \chi(T) \) is quite Pauli-like and does not show 
Curie-Weiss temperature dependence  
(see the inset of Fig. \ref{fig-chiVsT}) even
after a temperature independent term $\chi_0$ has been subtracted
to make 1/($\chi$-$\chi_0$) as linear as possible.  

Our  powder neutron diffraction measurements found magnetic peaks 
below $\sim$100~K, 
demonstrating that the 125~K features in $\chi$(T)  and $\rho_c(T)$ 
arise from magnetic ordering. 
The temperature dependence of $\chi$ and the hysteresis
measurement at 2~K indicate that this ordering is antiferromagnetic. 
Indeed, the magnetic peaks in the neutron diffraction 
were indexed to a propagation vector of 
$\vect{k}$=\( (\frac{1}{3}, \frac{1}{3} , 0 ) \) corresponding 
to a commensurate spin-density wave that 
triples each in-plane unit cell dimension.
Fitting the neutron spectra 
gives a moment on the Cr sites that varies between 
a maximum of of 2.2 and 
a minimum of 0.6 $\mu_B$. 
The moment on the iron sites is much weaker, and 
indeed within the noise it is consistent 
with an earlier 
M\"ossbauer spectroscopy
measurement \cite{Rancourt} which found 
very weak magnetic order on the iron sites 
at 4.2~K,  
with the magnetic moment estimated to be 
0.1 $\pm\ 0.03\ \mu_B$. 
A full treatment of the magnetic structure and symmetry analysis 
will be published separately.

\begin{figure}
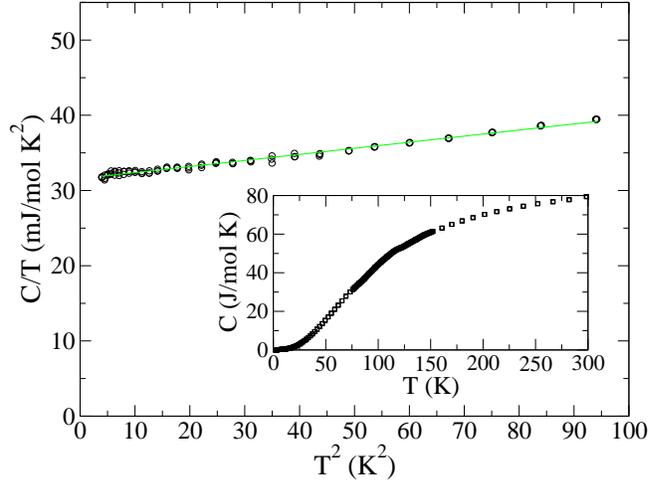
 
\onefigure[scale=0.34]{./Julian_Fig3.eps}
\caption{ C/T vs $T^2$ at low temperatures (circles). The green line is
a linear fit. The inset shows $C(T)$ from room temperature to 2~K. The N$\acute{e}$el transition produces a weak anomaly at 125~K.} 
\label{fig-CoverTvsT} 
\end{figure}

In striking contrast to the resistivity, the specific heat, shown in
Fig. \ref{fig-CoverTvsT},  exhibits classic Fermi liquid behaviour at low
temperatures: \( C(T)/T \) depends linearly on $T^2$.  
The Sommerfeld coefficient, $\gamma=C_{el}/T$,  extracted from the data in Fig. \ref{fig-CoverTvsT} 
gives a value of  $\rm {31.6\  mJ/mol~K^2}$, which  is quite high for a
$d$-electron system. 
We find that 
$\gamma$  varies between samples by up to 20\%, 
however 
the linear relationship of \( C(T)/T\)  vs $T^2$ holds in all samples  down
to at least 1~K.   
At high temperatures, the specific heat shows a weak anomaly 
near the N$\acute{e}$el temperature (see inset of Fig.\ \ref{fig-CoverTvsT}). 

To compare the enhancements of the specific heat
coefficient and the spin susceptibility, the Wilson ratio, $R_W =K \chi
/\gamma$, is commonly applied, where $K$ is a scale factor which gives a dimensionless value
$R_W=1$ for a free electron gas. For a Kondo system,
a value of $R_W=2$ is expected.  In FeCrAs, however, we find  $R_W \sim 5$
from the $c$-axis susceptibility and $R_W \sim 4$ from the $a$-axis
susceptibility.

\section{Discussion}

The unusual features of our results are as follows: 1) both the in-plane
and $c$-axis resistivities show strong non-metallic behaviour with 
negative ${\rm d} \rho/{\rm d} T $ 
to beyond $\sim$800~K, the highest temperatures we measured; 
2) at 
low temperatures, both the in-plane and  $c$-axis resistivities have
strongly non-metallic, non-Fermi-liquid behaviour, while the specific
heat is Fermi-liquid like; 
3) the linear coefficient of specific heat  $\gamma$ is
unusually large for a $d$-electron material; 
4) there is a large, roughly temperature independent (i.e.\ Pauli-like)
susceptibility, giving a Wilson ratio of between 4 and 5; 
and 
5) the ordered moment on the iron
sites is unusually small for an iron compound.

Some of these features are common to the parent compounds of the new
iron-pnictide superconductors, notably the large value of $\gamma$, the
high resistivity above the magnetic ordering transition, the weakness
of magnetism on the iron sites, the Pauli-like susceptibility  and 
the high Wilson ratio 
(see e.g.\ \cite{Kamihara,Ren,Rotter,Sasmal}).  
Even the magnetic ordering transition temperature is similar, 
however  
the magnetism here is of a very different origin: 
FeCrAs 
adopts a \( (\frac{1}{3}, \frac{1}{3} , 0 ) \) 
in-plane antiferromagnetic
structure but this ordering is clearly driven by the 
chromium sublattice, while in the layered 
pnictides the SDW-magnetic order is driven by the Fe sites.
This may explain why the  incoherent transport in the 
layered iron-pnictides disappears below \( T_{SDW} \), while 
in FeCrAs  it continues to \( T \rightarrow 0\)~K. (It should be 
noted that in polycrystalline 
LaFeOAs the resisitivity at low temperature does show an 
upturn as $ T \rightarrow 0$~K, which may be intrinsic or 
a grain boundary effect \cite{Kamihara}.)  

The similarities suggest that there may be physics 
that is common to the incoherent 
states of both FeCrAs and the layered iron-pnictides. 
For example,   
the fall in \( \rho_c(T) \) just below \( T_N \sim 125\)~K is similar to that for 
\( T < T_{SDW} \) in the layered materials, and may be a 
signature of a spin-fluctuation contribution to the resistivity. 
Dai et al.\ \cite{Dai08} have suggested that the incoherence of the 
layered iron-pnictides arises from spin-fluctuation scattering 
enhanced by frustrated exchange interactions. 
Similar arguments may apply in FeCrAs, 
and indeed the crystal structure 
of FeCrAs is more amenable to frustration than that of the layered 
iron-pnictides.

\begin{figure} 
\onefigure[scale=0.4]{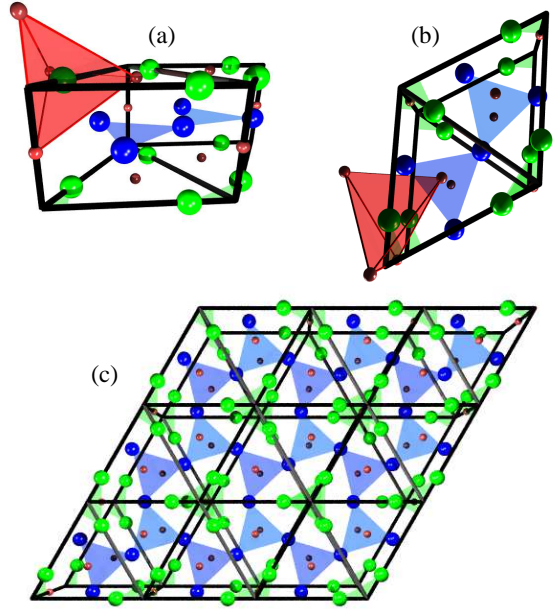}
\caption{Crystal structure of FeCrAs.  Arsenic atoms
are shown as small red balls,  Cr atoms are blue and Fe atoms are 
green.  (a) and (b): the primitive unit cell
from the side and the top respectively.  
The red tetrahedron shows the tetrahedral coordination of iron by arsenic. 
(c): looking down the $c$-axis at a three by three slice. 
The network of  blue triangles illustrates 
the distorted Kagome lattice on which the Cr atoms sit, 
while the green triangles illustrate the triangular lattice of Fe ``trimers".} \label{fig-xtalStructure}
\end{figure}

The $\rm {Fe_2P}$ crystal structure can be
viewed in a number of ways \cite{Fruchart}. Fig.\ \ref{fig-xtalStructure} (c)
emphasizes the  distinct chromium (blue atoms) and iron (green
atoms) sublattices. The Cr sublattice, which carries the bulk of
the magnetic order, can be viewed as a distorted
Kagome lattice, while  
the iron atoms form ``trimers'', the green triangles of
Fig. \ref{fig-xtalStructure}, 
that lie on a triangular lattice. 
These planes of iron trimers and the chromium Kagome planes are
alternately stacked along the $c$-axis, and the triangular network of
trimers in particular offers interesting possibilities for frustration. 
We note that  the small ordered Fe magnetic moment, the
very weak dependence of $\rho$(T), $\chi$(T) and  $C(T)/T$ on magnetic
field, and signatures of magnetic freezing in these crystallographically
well-ordered crystals, are all typical of magnetically frustrated systems
\cite{Nakatsuji,Saunders}.

However,
the antiferromagnetic order is a major
complication in this scenario, 
as is the absence of Curie-Weiss susceptibility. 
An even more serious difficulty is the energy scale: 
the related tetragonal materials \( \rm Fe_2As \) and \( \rm Cr_2As \) 
order antiferromagnetically at 350~K \cite{Katsuraki64} 
and 393~K \cite{Yuzuri}. 
If we take this as the 
typical scale of the exchange energy $J$ in these systems, 
it is then difficult to see how the non-metallic resistivity could 
persist to temperatures that are twice as high.
In most spin fluctuation systems the resistivity is a {\em rising} function of 
temperature, and it saturates when \(T\) is larger than $J$.
Finally, we note that the weak magnetism on the iron sites is probably 
due to electronic structure effects, not magnetic frustration. 
LMTO calculations by Ishida et al.\ \cite{Ishida} consistently find for 
the Fe$_2$P structure that the tetrahedrally coordinated site (the 
Fe site in FeCrAs) has a small moment, while the ``pyramidal" site, 
coordinated by five As atoms (the Cr site in FeCrAs),  
has a large moment. 
Indeed for FeCrAs they find that the density of states of the 
$3d$-bands of Fe is below the Stoner criterion, and although they 
found that a ferromagnetic ground state is energetically favoured, 
the antiferromagnetic state that 
they used for comparison was quite different from the one that we have found.

A very different explanation of the high temperature resistivity might be the 
local physics arising on the 
tetrahedrally coordinated Fe sites \cite{Haule,Shorikov,Laad,Sawatzky}.  
The bad semiconductor/bad metal picture of the incoherent metallic state 
of the layered iron-pnictides that arises in DMFT calculations 
is appealing because 
the energy scale is not the inter-site exchange coupling $J$. 
Rather it is the on-site inter-orbital Coulomb interaction, estimated 
to be on the order of a few eV \cite{Sawatzky}. 
We  therefore believe that FeCrAs may be an interesting system for 
exploring this new physics.  
We emphasize, however, that the strong temperature dependence of the  
non-metallic resistivity of FeCrAs is different from the roughly temperature 
independent resistivity seen at 
high temperatures in the parent compounds of the iron-pnictide superconductors.

Whether the DMFT picture explains the low temperature limit 
is not clear. 
Laad et al.\ \cite{Laad}, in attempting to explain optical conductivity, find 
a scattering rate that is sub-linear in \(\omega\), which may translate into 
a resistivity that is sub-linear in \(T\). 
However, the most striking feature of our low temperature results -- 
the incompatibility of the  non-metallic and non-Fermi-liquid resistivity
with the classic Fermi-liquid specific heat  -- 
has not been discussed within either the spin-fluctuation 
or the DMFT pictures.  

There are a number of materials that show strong non-metallic 
resistivity over a large temperature range, terminating at low 
temperature in  a non-Fermi-liquid state. 
Underdoped cuprates are an obvious example \cite{Ando}, 
however in that case the \( T \rightarrow 0\)~K limit of \( \rho(T) \) has a 
logarithmic divergence, not power-law behaviour. 

Some disordered heavy fermion systems \cite{Maple95,Stewart01} 
have non-metallic resistivity over a large temperature range, 
with low temperature linear-in-\(T\) behaviour,
however in all of these cases  
the specific heat 
also shows strong non-Fermi-liquid behaviour as \(T \rightarrow {\rm 0~K}\), typically \( C(T)/T \sim - \ln(T) \).
For example, in $\rm{CeRh_2Ga}$,  \( C(T)/T \) is logarithmic in \(T\) and 
rises from
$\sim$150 $\rm{mJ/mol~K^2}$ to $\sim$480 $\rm{mJ/mol~K^2}$ between
6~K  and  2~K \cite{Chen}. 
In contrast, the specific heat of FeCrAs is only weakly temperature dependent 
at low temperature and it is linear-in-$T$ as $T\rightarrow 0$~K.
Moreover, the disordered Kondo systems 
show dramatic
rises in $\chi(T)$ over the whole temperature range, typically by a factor of 
around 10 between room
temperature and  1~K, with increasing slope as T $\rightarrow$ 0~K.
The Pauli-like $\chi(T)$ of FeCrAs is completely different. 
We note additionally that the disordered Kondo systems 
have a large magnetoresistance 
at low temperature, while FeCrAs does not. 

It should be noted that 
there is one intriguing exception among the heavy fermion materials to the divergent \( C(T)/T\) 
accompanying a non-metallic non-Fermi-liquid resistivty, and that is another arsenide, 
$\rm{CeCuAs_2}$, in which  \( C(T)/T \) exhibits a fall, instead of 
logarithmic increase, below 2~K \cite{Sengupta}.  
However in this material too the susceptibility rises very strongly at low temperature,
unlike FeCrAs. The under-screened Kondo effect  has been suggested as an explanation of the non-metallic non-Fermi-liquid 
resistivity in $\rm {CeCuAs_2}$ \cite{Florens}, but given the apparent quenching of iron local moments by tetrahedral coordination 
with As \cite{Ishida,Sawatzky}, this scenario seems unlikely in FeCrAs.

Finally, disorder-induced localization effects, which have
been invoked to explain  non-metallic non-Fermi-liquid behaviour in the moderately
site-disordered U heavy fermion compounds  $\rm {URh_2Ge_2}$ \cite{Sullow} and
$\rm {UCu_4Pb}$ \cite{Otop} in the range of $\sim$1~K to 300~K, seem
to be ruled out because those materials exhibit quite a large 
magnetoresistance, clear signals in the Hall effect, 
and non-Fermi-liquid behaviours in heat capacity and magnetic susceptibility,  
while FeCrAs has none of these properties.

Thus we believe that the behaviour of FeCrAs is qualitatively different from 
any previous observations of non-metallic, non-Fermi-liquid behaviour 
in the \( T \rightarrow 0\)~K limit. 

It is natural to hypothesize that the linear low-temperature 
specific heat arises from 
fermionic excitations that are distinct from the non-metallic
non-Fermi-liquid charge carriers. 
Such ``neutral" fermions arise in some theories
of fractionalization in insulating, frustrated spin liquids
\cite{Lee,Affleck,Baskaran}, and may recently have been observed experimentally
 \cite{Kagawa}. 
Fractionalization has also been  introduced to  describe
antiferromagnetic quantum criticality in some heavy fermion compounds
\cite{Pepin,Senthil}.  
A key ingredient of most fractionalization scenarios
is a geometrically frustrated spin system (see e.g. \cite{Moessner}).
We have noted some difficulties with this picture in FeCrAs, 
but it may be relevant that 
the DMFT studies of the layered iron-pnictides 
\cite{Shorikov,Laad} find ``orbitally selective" 
effects that can be tuned by varying the on-site interactions. 
Although it is not believed that any of the $3d$-orbitals of Fe in the 
layered iron-pnictides are Mott localized, 
the local environment of the iron sites in FeCrAs may be sufficiently 
different from the layered systems 
that some of the $3d$-orbitals do localize, 
producing 
local degrees of freedom that do not contribute to 
charge transport. 
It is far from clear, however, if these degrees of freedom would be 
fermionic at low temperatures.


\section{Conclusion}
In summary, FeCrAs is an unusual $d$-electron system in which non-Fermi-liquid behaviour
has been observed. Its  resistivity shows  non-Fermi-liquid behaviour obeying
$\rho(T)=\rho_0-AT^{x}$ at low temperatures with \( x = 0.60 \pm 0.05\) for  
\( \rho_a \) and \(x = 0.70\pm 0.05 \) for \( \rho_c\), 
while at high
temperatures $\rho$ decreases with increasing
temperature up to at least 800~K. The low temperature specific heat,
on the other hand, exhibits typical Fermi-liquid behaviour with  a linear
temperature dependence. 
This low temperature state is  very robust in the sense that
it exhibits very weak dependence on field, and the same behaviour is 
observed in all the samples we have measured. 
The high temperature behaviour may be an extreme example 
of the ``bad semiconductor'' state found by DMFT calculations for the layered iron-pnictides, 
and 
although the low temperature state may have the same origin, it deserves 
further study as it has some features expected  of a fractionalized electron system  
which have not, to our knowlege, been observed before.

\acknowledgments

We are grateful to H.Y.\ Kee, J.\ Hopkinson, A.P.\ Mackenzie and Z.\ Tesanovic
for  valuable  discussions and  to Y.J. Kim  and F.\ Wang for assistance in our
magnetic hysteresis and susceptibility  measurements.  

\bibliography{FeCrAs_bibli}




\end{document}